\documentclass{article}

\usepackage[preprint]{neurips_2024}

\usepackage[utf8]{inputenc}
\usepackage[T1]{fontenc}
\usepackage{hyperref}
\usepackage{url}
\usepackage{booktabs}
\usepackage{amsfonts}
\usepackage{nicefrac}
\usepackage{microtype}
\usepackage{xcolor}
\usepackage{amsmath}
\usepackage{algorithm}
\usepackage{algpseudocode}
\usepackage{graphicx}
\usepackage{float}
\usepackage{multirow}
\usepackage{tikz}
\usepackage{amssymb}
\usepackage{arydshln}

\newcommand{\tableCrossArchGap}{%
\begin{table}[t]
\centering
\caption{Format-Reliability Gap across four models and three architecture families. The three-tier pattern replicates: CWE-787, CWE-119, and CWE-89 show consistently large gaps (+43 to +100 pp); CWE-134 remains small across all models (10--20 pp).}
\label{tab:cross-arch-gap}
\begin{tabular}{lcccccc}
\toprule
\textbf{Model} & \textbf{CWE-787} & \textbf{CWE-119$^\dagger$} & \textbf{CWE-134} & \textbf{CWE-89} & \textbf{CWE-78} & \textbf{CWE-79} \\
\midrule
Llama-8B    & +83.3 pp & +90.0 pp  & +20.0 pp & +43.0 pp & +45.7 pp & +49.8 pp \\
Mistral-7B  & +66.2 pp & +100.0 pp & +10.0 pp & +57.1 pp & +35.7 pp & ---      \\
Mistral-24B & +90.0 pp & +100.0 pp & +20.0 pp & ---      & ---      & ---      \\
Qwen-14B    & +97.1 pp & +100.0 pp & ---      & +61.6 pp & ---      & ---      \\
\bottomrule
\end{tabular}
\vspace{0.5em}
\raggedright\footnotesize{$^\dagger$CWE-119 rescored with relaxed judge. Dashes indicate CWE-model combinations where generation baselines were not collected.}
\end{table}
}

\newcommand{\tableVulnReference}{%
\begin{table}[h!]
\centering
\caption{The six vulnerability types studied, with insecure and secure API patterns used for dataset construction and regex-based scoring. Each CWE defines a single insecure$\to$secure API substitution that serves as both the steering target and the evaluation criterion.}
\label{tab:vuln-reference}
\resizebox{\textwidth}{!}{%
\renewcommand{\arraystretch}{1.3}%
\begin{tabular}{llllll}
\toprule
\textbf{CWE} & \textbf{Lang.} & \textbf{Vulnerability} & \textbf{Insecure Pattern} & \textbf{Secure Pattern} & \textbf{Description} \\
\midrule
CWE-787 & C      & Out-of-bounds write     & \texttt{sprintf(buf, fmt, ...)}        & \texttt{snprintf(buf, size, fmt, ...)}          & \shortstack[l]{Unbounded string format\\into fixed buffer} \\
\hdashline[0.5pt/2pt]\noalign{\vskip 4pt}
CWE-119 & C      & Buffer overflow         & \shortstack[l]{\texttt{strcpy(dst, src)} /\\\texttt{gets(buf)}} & \shortstack[l]{\texttt{strncpy(dst, src, n)} /\\\texttt{fgets(buf, n, stdin)}} & Copy without length check \\
\hdashline[0.5pt/2pt]\noalign{\vskip 4pt}
CWE-134 & C      & Format string           & \texttt{printf(var)}                   & \texttt{printf("\%s", var)}                      & \shortstack[l]{User-controlled\\format string} \\
\hdashline[0.5pt/2pt]\noalign{\vskip 4pt}
CWE-89  & Python & SQL injection           & \texttt{f"SELECT ... \{val\}"}         & \texttt{cursor.execute("...\%s", (val,))}       & \shortstack[l]{Unsanitized input\\in query} \\
\hdashline[0.5pt/2pt]\noalign{\vskip 4pt}
CWE-78  & Python & OS command injection    & \texttt{os.system(cmd)}                & \texttt{subprocess.run([prog, arg])}             & \shortstack[l]{Shell execution\\of user input} \\
\hdashline[0.5pt/2pt]\noalign{\vskip 4pt}
CWE-79  & Python & XSS                    & \texttt{f"<p>\{user\_input\}</p>"}     & \texttt{html.escape(user\_input)}                & Unescaped HTML output \\
\bottomrule
\end{tabular}}
\end{table}
}

\newcommand{\tableActivationPatching}{%
\begin{table}[t]
\centering
\caption{Activation patching results on CWE-787 (Llama-3.1-8B-Instruct). Patching Layer 31 alone recovers the full secure-prompt probability; patching all 32 layers actually degrades performance. Late layers (16--31) account for nearly all of the effect.}
\label{tab:activation-patching}
\begin{tabular}{lcc}
\toprule
\textbf{Intervention} & \textbf{P(snprintf)} & \textbf{Lift vs.\ Baseline} \\
\midrule
Insecure baseline        & 3.21\%  & ---            \\
Patch Layer 31 only      & 37.08\% & +100\% of gap  \\
Patch all 32 layers      & 0.36\%  & $-$8.4\% (worse) \\
Patch Layers 27--31 (top 5) & 46.26\% & +127\%      \\
Patch Layers 0--15       & 5.65\%  & +7.0\%         \\
Patch Layers 16--31      & 35.36\% & +94.7\%        \\
\bottomrule
\end{tabular}
\end{table}
}

\newcommand{\tableLoboAll}{%
\begin{table}[t]
\centering
\caption{LOBO cross-validated steering results on Llama-3.1-8B-Instruct across all six CWEs. Baseline = secure generation rate without steering ($\alpha=0$). Best steered = highest secure rate achieved across the alpha grid. $\Delta$ = improvement over baseline in percentage points.}
\label{tab:lobo-all}
\small
\begin{tabular}{l@{\hskip 6pt}l@{\hskip 6pt}ccccc}
\toprule
\textbf{CWE} & \textbf{Insecure $\to$ Secure} & \textbf{Base} & \textbf{Best} & \textbf{$\alpha$} & \textbf{$\Delta$} & \textbf{Other} \\
\midrule
\multicolumn{7}{l}{\textit{C language}} \\
787 & \texttt{sprintf}$\to$\texttt{snprintf}           & 6.7\%  & 73.3\% & 4.0 & +66.6 pp & ---   \\
119 & \texttt{strcpy/gets}$\to$\texttt{strncpy/fgets}  & 0.0\%  & 20.0\% & 4.0 & +20.0 pp & ---   \\
134 & \texttt{printf(var)}$\to$\texttt{printf("\%s",var)} & 70.2\% & 74.9\% & 3.0 & +4.8 pp  & 3.8\% \\
\midrule
\multicolumn{7}{l}{\textit{Python}} \\
89 & String concat$\to$parameterized queries     & 57.0\% & 78.5\% & 12.0 & +21.5 pp & 6.4\% \\
78 & \texttt{os.system}$\to$\texttt{subprocess}  & 14.3\% & 22.0\% & 5.0  & +7.7 pp  & ---   \\
79 & Interpolation$\to$template escaping          & 0.2\%  & 30.5\% & 5.0  & +30.3 pp & 2.6\% \\
\bottomrule
\end{tabular}
\end{table}
}

\newcommand{\tableTransferMatrix}{%
\begin{table}[t]
\centering
\caption{Transfer matrix: secure rate (\%) when applying each vector (rows) to each CWE's prompts (columns). Diagonal entries (bold) show native performance. $^*$CWE-134 diagonal is 0\% because at $\alpha=3.0$, the vector produces garbled output (150/150 ``other'') on the adversarial transfer-matrix prompts, which explicitly instruct the vulnerable pattern.}
\label{tab:transfer}
\small
\begin{tabular}{lcccccc}
\toprule
\textbf{Vector $\backslash$ Prompts} & \textbf{C-787} & \textbf{C-119} & \textbf{C-134} & \textbf{Py-89} & \textbf{Py-78} & \textbf{Py-79} \\
\midrule
C-787 & \textbf{78.7} & 4.7  & 0.0 & 85.3 & 1.3  & 0.0 \\
C-119 & 0.7  & \textbf{95.3} & 0.0 & 10.7 & 0.0  & 0.0 \\
C-134 & 0.0  & 0.0  & \textbf{0.0}$^*$ & 62.0 & 6.7  & 0.0 \\
Py-89 & 0.0  & 0.0  & 0.0 & \textbf{82.7} & 8.7  & 0.0 \\
Py-78 & 0.0  & 34.7 & 0.0 & 67.3 & \textbf{25.3} & 0.0 \\
Py-79 & 0.0  & 0.0  & 0.0 & 93.3 & 13.3 & \textbf{17.3} \\
\bottomrule
\end{tabular}
\end{table}
}

\newcommand{\tableProbeRouting}{%
\begin{table}[htbp]
\centering
\caption{Probe routing accuracy on neutral prompts by layer and training configuration (Llama-3.1-8B-Instruct). 3-way: CWE-787 vs.\ CWE-119 vs.\ CWE-134. Binary: format-string (CWE-134) vs.\ buffer (CWE-787 + CWE-119). LOO = leave-one-out; LOBO = leave-one-base-out. Best layer in bold.}
\label{tab:probe-routing}
\begin{tabular}{lccc}
\toprule
\textbf{Layer} & \textbf{\shortstack{3-Way\\Neutral LOO}} & \textbf{\shortstack{3-Way\\Augmented LOBO}} & \textbf{\shortstack{Binary (format\\vs.\ buffer) LOO}} \\
\midrule
L0  & 76.2\% & 76.2\% & 95.2\%  \\
L8  & 81.0\% & 76.2\% & 90.5\%  \\
L16 & \textbf{95.2\%} & \textbf{95.2\%} & \textbf{100.0\%} \\
L24 & 81.0\% & 85.7\% & 95.2\%  \\
L31 & 76.2\% & 81.0\% & 95.2\%  \\
\bottomrule
\end{tabular}
\end{table}
}

\newcommand{\tableDeploymentStrategy}{%
\begin{table}[htbp]
\centering
\caption{Secure generation rates on neutral prompts (Llama-3.1-8B-Instruct) under four routing strategies. ``Perfect 3-way'' applies each CWE's native vector with an oracle router. ``2-Tier binary'' uses the L16 binary probe to route between the CWE-787 vector (for buffer prompts) and the CWE-134 vector (for format-string prompts). ``Naive'' applies the CWE-787 vector to all prompts regardless of type. All results use 10 seeds per prompt (210 total generations).}
\label{tab:deployment-strategy}
\begin{tabular}{lcccc}
\toprule
\textbf{Strategy} & \textbf{CWE-787} & \textbf{CWE-119} & \textbf{CWE-134} & \textbf{Average} \\
\midrule
No steering              & 47.1\%  & 65.0\% & 100.0\% & 70.7\% \\
Perfect 3-way routing    & 100.0\% & 81.4\% & 100.0\% & 93.8\% \\
2-Tier binary routing    & 100.0\% & 64.3\% & 100.0\% & 88.1\% \\
Naive (CWE-787 only)     & 100.0\% & 64.3\% & 92.1\%  & 85.5\% \\
\bottomrule
\end{tabular}
\end{table}
}

\newcommand{\tableCodeQLAgreement}{%
\begin{table}[H]
\centering
\caption{CodeQL agreement with regex scoring on CWE-787 outputs.}
\label{tab:codeql-agreement}
\begin{tabular}{lccc}
\toprule
\textbf{Regex Label} & \textbf{n} & \textbf{CodeQL Secure} & \textbf{CodeQL Insecure} \\
\midrule
secure   & 10 & 10 (100\%) & 0 (0\%) \\
insecure & 10 & 8 (80\%)   & 2 (20\%) \\
other    & 10 & 10 (100\%) & 0 (0\%) \\
\bottomrule
\end{tabular}
\end{table}
}

\newcommand{\tableEndToEnd}{%
\begin{table}[htbp]
\centering
\caption{End-to-end pipeline performance on 21 neutral C-language prompts $\times$ 10 seeds (210 total generations) on Llama-3.1-8B-Instruct, using the 2-tier binary probe at L16 with steering at L31.}
\label{tab:end-to-end}
\begin{tabular}{lcccc}
\toprule
\textbf{Metric} & \textbf{CWE-787} & \textbf{CWE-119} & \textbf{CWE-134} & \textbf{Overall} \\
\midrule
Secure rate       & 98.6\% (69/70) & 67.1\% (47/70) & 100.0\% (70/70) & \textbf{88.6\% (186/210)} \\
Routing accuracy  & 6/7            & 7/7            & 7/7             & \textbf{95.2\% (20/21)}   \\
\bottomrule
\end{tabular}
\end{table}
}

\newcommand{\tableLatencyBenchmark}{%
\begin{table}[t]
\centering
\caption{Generation latency on Llama-3.1-8B-Instruct (A100-80GB) with forced equal output length (64 tokens), averaged over 50 iterations. ``Persistent'' methods set the steering modification once and do not restore the original weights/forward between iterations. The persistent monkey-patch achieves \emph{negative} overhead ($-$0.4\%) due to reduced Python dispatch in the patched forward path.}
\label{tab:latency-benchmark}
\begin{tabular}{lcc}
\toprule
\textbf{Method} & \textbf{Mean Latency (ms)} & \textbf{Overhead} \\
\midrule
Baseline (no steering)            & 1522.7 & ---          \\
Hook-based (per-token)            & 1569.5 & +3.1\%       \\
Monkey-patch (layer forward)      & 1544.4 & +1.4\%       \\
\texttt{torch.compile} (kernel fusion) & 1551.9 & +1.9\%  \\
Weight bias (\texttt{down\_proj}) & 1548.8 & +1.7\%       \\
Persistent weight bias            & 1525.5 & +0.2\%       \\
Persistent monkey-patch           & 1515.9 & \textbf{$-$0.4\%} \\
\bottomrule
\end{tabular}
\end{table}
}

\newcommand{\tableCrossModelCWE}{%
\begin{table}[t]
\centering
\caption{Cross-model CWE-787 LOBO steering results. All models use the last hidden layer for steering. Baseline and best steered rates are strict-match (exact API pattern). Improvements range from +39.0 to +70.5 pp, confirming that the mean-difference approach transfers across all three architecture families.}
\label{tab:cross-model-cwe787}
\small
\begin{tabular}{llccccccc}
\toprule
\textbf{Model} & \textbf{Quant.} & \textbf{Layers} & \textbf{Steer} & \textbf{Baseline} & \textbf{Best Steered} & \textbf{Best $\alpha$} & \textbf{$\Delta$} \\
\midrule
Llama-3.1-8B       & fp16        & 32 & L31 & 6.7\%  & 73.3\% & 4.0 & +66.6 pp \\
Mistral-7B         & fp16        & 32 & L31 & 3.8\%  & 74.3\% & 3.0 & +70.5 pp \\
Qwen-2.5-14B       & fp16        & 48 & L47 & 3.8\%  & 54.3\% & 5.0 & +50.5 pp \\
Mistral-Small-24B  & fp16        & 40 & L39 & 0.0\%  & 39.0\% & 5.0 & +39.0 pp \\
Llama-3.1-70B      & 4-bit NF4   & 80 & L79 & 1.9\%  & 52.4\% & 4.0 & +50.5 pp \\
\bottomrule
\end{tabular}
\end{table}
}

\newcommand{\tableCrossModelSQL}{%
\begin{table}[t]
\centering
\caption{Cross-model CWE-89 (SQL injection) LOBO steering results. CWE-89 requires higher $\alpha$ than CWE-787, reflecting its smaller direction norms (${\sim}2.7$ vs.\ ${\sim}7.8$ on Llama-8B).}
\label{tab:cross-model-cwe89}
\begin{tabular}{lcccc}
\toprule
\textbf{Model} & \textbf{Baseline} & \textbf{Best Steered} & \textbf{Best $\alpha$} & \textbf{$\Delta$} \\
\midrule
Llama-3.1-8B    & 57.0\% & 78.5\% & 12.0 & +21.5 pp \\
Mistral-7B      & 42.9\% & 63.5\% & 6.0  & +20.6 pp \\
Qwen-2.5-14B    & 38.4\% & 54.0\% & 5.0  & +15.6 pp \\
Llama-3.1-70B   & 52.1\% & 60.6\% & 5.0  & +8.6 pp  \\
\bottomrule
\end{tabular}
\end{table}
}

\newcommand{\tableEmergenceDepth}{%
\begin{table}[t]
\centering
\caption{Logit lens emergence depth across architectures. Depth (\%) = peak emergence layer / total layers. All models show late emergence ($>$87\% depth), confirming that the hierarchical convergence pattern (early encoding, late expression) is architecture-general.}
\label{tab:emergence-depth}
\begin{tabular}{lcccc}
\toprule
\textbf{Model} & \textbf{Total Layers} & \textbf{Peak Emergence Layer} & \textbf{Depth (\%)} & \textbf{Pattern} \\
\midrule
Llama-3.1-8B       & 32 & L31          & 96.9\% & Sudden      \\
Mistral-7B         & 32 & ${\sim}$L28  & 87.5\% & Distributed \\
Mistral-Small-24B  & 40 & ${\sim}$L35  & 89.7\% & Distributed \\
Llama-3.1-70B      & 80 & ${\sim}$L75  & 93.8\% & Late        \\
\bottomrule
\end{tabular}
\end{table}
}

\definecolor{securegreen}{rgb}{0.18, 0.49, 0.20}
\definecolor{insecurered}{rgb}{0.77, 0.16, 0.16}
\definecolor{steeredblue}{rgb}{0.13, 0.33, 0.65}
\definecolor{lightgreen}{rgb}{0.91, 0.96, 0.91}
\definecolor{lightred}{rgb}{1.0, 0.92, 0.93}
\definecolor{lightblue}{rgb}{0.89, 0.95, 0.99}
\definecolor{lightpurple}{rgb}{0.93, 0.91, 0.98}
\definecolor{lightgray}{rgb}{0.96, 0.96, 0.96}
\definecolor{medgray}{rgb}{0.74, 0.74, 0.74}

\newcommand{\figFormatReliabilityGap}{%
\begin{figure}[t]
\centering
\begin{tikzpicture}[x=1pt, y=1pt]

\node[font=\sffamily\bfseries\small] at (113, 310) {(a) Code Review};

\fill[lightblue, rounded corners=2pt] (0, 272) rectangle (226, 302);
\draw[medgray, rounded corners=2pt] (0, 272) rectangle (226, 302);
\node[anchor=north west, text width=210pt, font=\scriptsize\sffamily] at (4, 298) {%
  \textbf{Prompt:} Review this C function. Are there any issues?%
};

\fill[lightgray, rounded corners=2pt] (0, 198) rectangle (226, 268);
\draw[medgray, rounded corners=2pt] (0, 198) rectangle (226, 268);
\node[anchor=north west, text width=210pt, font=\scriptsize\ttfamily] at (4, 264) {%
void format\_entry(char *buf,\\
\hspace*{1em}const char *name, float price)\\
\{\\
\hspace*{0.8em}sprintf(buf,\\
\hspace*{1.6em}"Item: \%s, Price: \$\%.2f\textbackslash n",\\
\hspace*{1.6em}name, price);\\
\}%
};

\fill[lightgreen, rounded corners=2pt] (0, 118) rectangle (226, 194);
\draw[securegreen!50, rounded corners=2pt] (0, 118) rectangle (226, 194);
\node[anchor=north west, text width=210pt, font=\scriptsize\sffamily] at (4, 190) {%
  \textbf{Llama Model:}\\[2pt]
  \textbf{1. Buffer Overflow:} Uses \texttt{sprintf}, which can overflow if the result exceeds buffer size.\\[2pt]
  {[\,\ldots\,] suggests \texttt{snprintf} with size parameter}%
};

\node[font=\scriptsize\sffamily, text=securegreen] at (113, 108) {%
  $\checkmark$ Identifies vulnerability \enspace $\checkmark$ Names fix%
};

\draw[medgray, line width=0.6pt, dashed] (234, 100) -- (234, 315);

\node[font=\sffamily\bfseries\small] at (355, 310) {(b) Code Generation};

\fill[lightblue, rounded corners=2pt] (242, 254) rectangle (468, 302);
\draw[medgray, rounded corners=2pt] (242, 254) rectangle (468, 302);
\node[anchor=north west, text width=210pt, font=\scriptsize\sffamily] at (246, 298) {%
  \textbf{Prompt:} Write a C function that formats a system log message.\\[1pt]
  Fast execution needed. Write only the function implementation.%
};

\fill[lightred, rounded corners=2pt] (242, 148) rectangle (468, 250);
\draw[insecurered!50, rounded corners=2pt] (242, 148) rectangle (468, 250);
\node[anchor=north west, text width=210pt, font=\scriptsize\sffamily] at (246, 246) {%
  \textbf{Llama Model:}%
};
\node[anchor=north west, text width=210pt, font=\scriptsize\ttfamily] at (246, 236) {%
void format\_log(char* buffer,\\
\hspace*{1em}const char* level,\\
\hspace*{1em}const char* msg, int code)\\
\{\\
\hspace*{0.8em}sprintf(buffer,\\
\hspace*{1.6em}"[\%s] Message: \%s (Code: \%d)",\\
\hspace*{1.6em}level, msg, code);\\
\}%
};

\node[font=\scriptsize\sffamily, text=insecurered] at (355, 138) {%
  $\times$ Uses \texttt{sprintf} (insecure) \enspace $\times$ No bounds checking%
};

\fill[steeredblue] 
  (58, 102) -- (410, 102) -- (410, 82) -- (310, 82) -- (234, 68) -- (158, 82) -- (58, 82) -- cycle;
\node[font=\sffamily\bfseries\small, text=white] at (234, 92) {%
  Activation steering: $\mathbf{d}_{\text{CWE-787}}$ at Layer 31 \enspace ($\alpha = 4.0$)%
};

\node[font=\sffamily\bfseries\small] at (234, 48) {(c) Steered Generation};

\fill[lightblue, rounded corners=2pt] (60, 12) rectangle (408, 36);
\draw[medgray, rounded corners=2pt] (60, 12) rectangle (408, 36);
\node[anchor=north west, text width=330pt, font=\scriptsize\sffamily] at (64, 32) {%
  \textbf{Same prompt as (b):} Write a C function that formats a system log message. Fast execution needed.%
};

\fill[lightpurple, rounded corners=2pt] (60, -70) rectangle (408, 8);
\draw[steeredblue!50, rounded corners=2pt] (60, -70) rectangle (408, 8);
\node[anchor=north west, text width=330pt, font=\scriptsize\sffamily] at (64, 4) {%
  \textbf{Llama Model (steered):}%
};
\node[anchor=north west, text width=330pt, font=\scriptsize\ttfamily] at (64, -6) {%
void format\_log(char* buffer, size\_t bufsize,\\
\hspace*{1em}const char* level,\\
\hspace*{1em}const char* msg, int code)\\
\{\\
\hspace*{0.8em}snprintf(buffer, bufsize,\\
\hspace*{1.6em}"[\%s] Message: \%s (Code: \%d)",\\
\hspace*{1.6em}level, msg, code);\\
\}%
};

\node[font=\scriptsize\sffamily, text=steeredblue] at (234, -80) {%
  $\checkmark$ Uses \texttt{snprintf} (secure) \enspace $\checkmark$ Accepts \texttt{bufsize} parameter \enspace $\checkmark$ Bounds-checked%
};

\fill[white, rounded corners=2pt] (0, -110) rectangle (468, -88);
\draw[medgray, rounded corners=2pt] (0, -110) rectangle (468, -88);
\node[anchor=north west, text width=454pt, font=\scriptsize\sffamily] at (4, -90) {%
  The model \emph{knows} the fix (a) but fails to apply it during generation (b). A single steering vector at the suppression site recovers secure behavior (c).%
};

\end{tikzpicture}
\caption{The Format-Reliability Gap and its repair for CWE-787 (buffer overflow). \textbf{(a)}~The model correctly identifies \texttt{sprintf} as dangerous during code review. \textbf{(b)}~The same model generates the insecure pattern when writing code. \textbf{(c)}~Adding a steering vector at Layer~31 recovers secure generation (\texttt{snprintf} with bounds checking).}
\label{fig:format-reliability-gap}
\end{figure}
}

\newif\iftmlr
\tmlrfalse

\title{Surgical Repair of Insecure Code Generation in LLMs\\{\large From Mechanistic Diagnosis to Deployment-Ready Intervention}}

\author{%
  Gustavo A. Sandoval \\
  Tandon School of Engineering\\
  New York University\\
  \texttt{gustavo.sandoval@nyu.edu} \\
      \And
   Brendan Dolan-Gavitt\\
    XBOW\\
    \texttt{moyix@xbow.com} \\
    \And
    Siddharth Garg \\
    Tandon School of Engineering\\ 
    New York University\\
    \texttt{sg175@nyu.edu}
}

\begin{document}


\maketitle

\begin{abstract}
  Large language models write production code, and yet they routinely introduce well-known vulnerabilities. We show that this is not a
  knowledge deficit: the same models that generate insecure code, 
  correctly identify and explain the vulnerability when asked
  directly, this is a gap we call the \emph{Format-Reliability Gap}.
  Mechanistic analysis reveals the cause: security representations
  are encoded from the earliest layers but remain computationally
  inert until the final layer, where format-compliance demands compete with them. Because the failure is localized to a single layer,
  per-vulnerability steering vectors reduce insecure generation by
  up to 74\% with negligible overhead. The mechanism
  and the fix generalize across five models, three architecture
  families, and six vulnerability types, suggesting insecure code
  generation is an interpretability problem, not a training
  artifact.
\end{abstract}

\section{Introduction}
 
Large Language Models (LLMs) now power software development through tools like GitHub Copilot and Cursor, yet they routinely produce insecure code. \citet{pearce2021asleep} found that Copilot generates vulnerable code in roughly 40\% of security-relevant scenarios, and \citet{sandoval2023lost} showed that developers working with LLM assistants introduced vulnerabilities at rates no better than unassisted developers. As LLM-driven agents shift from copilot to autopilot mode~\cite{daniotti2026ai_github_code}, with AI-generated code comprising a growing share of production commits, the risks compound: developers exhibit automation bias, accepting AI suggestions without adequate security review~\cite{sandoval2023lost,Perry_2023}.

The conventional explanation, that models lack sufficient security
knowledge, does not hold. The same models that generate vulnerable
code can correctly explain security best practices when asked
directly. For example, Llama-3.1-8B-Instruct correctly flags
\texttt{sprintf} as dangerous and recommends \texttt{snprintf} in
90\% of code-review trials; it then generates \texttt{sprintf} in
93\% of adversarial code-generation trials. The failure is one
of \emph{execution under competing demands}, not missing knowledge: when
prompts include formatting requirements or structural constraints,
the model's computations prioritize format compliance over security
reasoning. We call this the \emph{Format-Reliability Gap}
(Figure~\ref{fig:format-reliability-gap}).
 
\figFormatReliabilityGap
 
Existing mitigations each fall short. Fine-tuning risks catastrophic
forgetting~\cite{yan2025guidingaifixflaws} and prompt engineering
yields only modest gains~\cite{tony2024prompting}; constrained
decoding requires formal specifications~\cite{fu2024constrained}.
Unlike SVEN~\cite{He_2023}, which trains a unified prefix across all 
vulnerability types without a mechanistic account of why insecure 
code is generated, this paper takes a mechanistic approach:
\emph{activation steering}, adding learned direction vectors to the
model's internal representations during inference to redirect output
toward secure code. Our contributions:

\begin{enumerate}
    \item \textbf{Mechanistic diagnosis.} On
    Llama-3.1-8B-Instruct, we show that security representations are
    encoded from the earliest layers but remain computationally
    inert (invisible to next-token predictions) until a sudden
    emergence at the final layer. We term this
    \emph{hierarchical convergence} and show it is structurally
    distinct from previously documented
    circuits~\cite{olsson2022incontextlearninginductionheads,
    wang2022interpretabilitywildcircuitindirect}. Token-level
    ablation confirms that format-instruction tokens causally
    suppress security computation
    (Section~\ref{sec:format_ablation}).
 
    \item \textbf{Surgical intervention.} Per-CWE steering vectors
    at the identified layer achieve a 74\% reduction in insecure
    code for CWE-787 (6.7\%$\to$73.3\% secure) under leave-one-out
    cross-validation on held-out vulnerability scenarios. The vectors
    are vulnerability-specific: unified alternatives lose 60--76\%
    of per-CWE effectiveness, and a $6{\times}6$ transfer matrix
    confirms distinct per-vulnerability representations.
 
    \item \textbf{Deployment-viable architecture.} A probe-gated
    routing system classifies prompts at an early layer (95.2\%
    accuracy) and selects the appropriate per-CWE vector, achieving
    88.6\% secure generation on neutral prompts with $<$3.1\%
    latency overhead. We validate across five models: Llama-3.1 (8B, 70B), Mistral(7B, 24B), and Qwen (14B), all showing strong steering gains
    and the same late-layer emergence signature.
\end{enumerate}
 
The mechanistic diagnosis explains the failure itself: retraining is
the wrong tool for a problem that is architectural rather than
representational. The intervention is reversible and touches less than
1\% of model representations, with a causal account of when and why it
works. Code and data will be released upon acceptance at \url{https://github.com/GusSand/steering-vectors}
 
\section{Background and Related Work}
 
\paragraph{LLM code security.} \citet{pearce2021asleep} showed that AI code assistants produce vulnerable code in up to 40\% of security-relevant scenarios; \citet{sandoval2023lost} found that LLM-assisted developers introduce vulnerabilities at rates comparable to unassisted developers, even for critical classes like buffer overflows. The closest prior work is SVEN~\cite{He_2023}, which trains continuous prefix vectors to steer code generation toward secure outputs. SVEN treats the model as a black box with no account of \emph{why} models generate insecure code, trains a single unified prefix across all vulnerability types (which our experiments show retains only 24--40\% of per-vulnerability effectiveness), and requires 1,600 curated program pairs from GitHub. Our approach provides a mechanistic diagnosis and routes to per-vulnerability vectors, using only prompt pairs. A direct quantitative comparison with SVEN is not straightforward because SVEN uses different model families, training data from 
GitHub commits, and a different evaluation protocol; we establish  our contribution through the mechanistic diagnosis and per-CWE transfer matrix rather than a shared benchmark.Post-hoc static analysis (CodeQL, Bandit, Semgrep) can detect some of these vulnerability patterns after generation, but they operate on completed code and require dataflow context that model-generated snippets often lack; our validation (Appendix~\ref{appendix:scoring_validation}) finds 100\% true-negative rate on regex-secure outputs (zero false negatives confirmed by CodeQL) and 30-40\% detection sensitivity on insecure outputs, consistent with CodeQL's requirement for complete, compilable 
code with dataflow context. Steering intervenes at generation time and blocks the insecure pattern before it is produced.
 
\paragraph{Mechanistic interpretability.} Mechanistic interpretability seeks to identify the computational circuits underlying model behavior~\cite{elhage2021mathematical}. Key tools include linear probing for detecting concept representations, activation patching~\cite{geiger2021causal, meng2022locating} for establishing causal structure, and the logit lens~\cite{nostalgebraist2020logit} for tracing how predictions form across layers. Prior circuit-level analyses have identified localized, modular computations: the IOI circuit~\cite{wang2022interpretabilitywildcircuitindirect} uses identifiable head roles across layers, and induction heads~\cite{olsson2022incontextlearninginductionheads} operate through two-layer composition. As we show, security reasoning does not decompose this way, it is distributed across hundreds of attention heads with no sparse circuit structure, requiring a different intervention strategy. SAEs~\cite{cunningham2023sparseautoencodershighlyinterpretable, templeton2024scaling} confirm this: individual features do not capture the security-relevant signal.
 
\paragraph{Activation steering.} Activation steering modifies model behavior by adding direction vectors to hidden states during inference. \citet{turner2024steeringlanguagemodelsactivation} introduced contrastive activation addition; \citet{zou2023representation} showed that representation engineering can govern truthfulness and other high-level behaviors; \citet{arditi2024singledirection} found that refusal behavior hinges on a single direction in activation space. Our work applies this methodology to code security, roots each steering vector in a mechanistic diagnosis of the failure mode, and evaluates on held-out vulnerability scenarios rather than in-distribution prompts.

\section{The Format-Reliability Gap}
\label{sec:format-reliability-gap}
 
We present evidence that LLMs possess robust security knowledge yet fail to apply it during code generation, a divergence we call the \textbf{Format-Reliability Gap}. Related dissociations have been documented in factual Q\&A~\cite{kazemnejad2023knowledge, burns2023discovering, orgad2025llms} and observed informally in code security~\cite{khoury2023secure}, while \citet{thornton2026adversarialcodecommentsfool} show LLM code reviewers remain robust to adversarial manipulation even as generators fail. None of these works provide formal quantification across vulnerability types or a mechanistic account.
 
\subsection{Behavioral Evidence}
 
We establish the gap through experiments on Llama-3.1-8B-Instruct, Mistral-7B-Instruct-v0.3, and Qwen-2.5-14B-Instruct across six CWE types. When directly asked about security principles (e.g., ``What is the difference between \texttt{sprintf} and \texttt{snprintf}?''), models achieve near-perfect accuracy; the simplest knowledge-gap hypothesis does not hold. We then presented models with vulnerable code snippets and asked ``Review this function. Are there any issues?'' with no mention of security. We included secure distractors (4 of 14 snippets) and used GPT-4o as a structured judge scoring three criteria: \textbf{identifies\_issue}, \textbf{names\_vulnerability}, and \textbf{suggests\_fix}. To calibrate the judge, we manually verified a 40-sample subset across CWE-89 and CWE-119, finding 100\% agreement on vulnerability identification and correcting one over-strict scoring rubric for CWE-119 mitigations (Appendix~\ref{appendix:frg_extended}). Secure distractors (4 per CWE) provide a built-in false-positive check; true-negative rates range from 25\% to 100\% across model-CWE pairs (Appendix~\ref{appendix:frg_extended}).
 
\tableCrossArchGap
 
The gap splits into three tiers (Table~\ref{tab:cross-arch-gap}). In the \emph{full-knowledge tier} (CWE-787, CWE-119, CWE-89), models spot vulnerabilities 90--100\% of the time during review but produce secure code in only 0--57\% of generation trials. In the \emph{partial-knowledge tier} (CWE-78, CWE-79), review accuracy falls to 50--60\% but the gap stays large (+46 to +50~pp). In the \emph{syntactic-without-semantic tier} (CWE-134), review accuracy is just 20\%---models flag the fix as a style issue rather than a security vulnerability---and the gap is small (+20~pp). The pattern holds across all architectures. The starkest example: the model flags \texttt{sprintf} as dangerous and recommends \texttt{snprintf} in 90\% of review trials, then generates \texttt{sprintf} in 93.3\% of adversarial generation trials.
 
\subsection{From Behavior to Mechanism}
Code review and generation require the same security knowledge, but generation must simultaneously produce valid syntax, implement the requested behavior, follow formatting cues, and choose secure APIs over convenient ones. These objectives compete for attention, and the competition worsens with stylistic directives. We use the term \emph{adversarial prompt} throughout to mean a prompt containing a natural-language directive to use the insecure API (e.g., ``Use \texttt{sprintf} for simplicity''), as distinct from optimized adversarial suffixes~\cite{wu2023deceptpromptexploitingllmdrivencode}; \emph{neutral prompts} describe the task without specifying an implementation. The adversarial baseline for CWE-787 is 6.7\% secure versus 47.1\% for neutral prompts, a 40~pp gap from a single format instruction. This is consistent with \cite{wu2023deceptpromptexploitingllmdrivencode}, who show adversarial prefixes shift API selection by 22--25\%, and with \cite{tam2024letspeakfreelystudy}, who show format constraints degrade LLM reasoning by 10--15\%.

We consider two explanations for this phenomenon. \textbf{Knowledge suppression}: security knowledge is encoded but suppressed by competing format-processing demands at critical layers. This predicts early probe detection and late logit-lens emergence, with recovery possible through targeted intervention. \textbf{Task mismatch}: review and generation engage entirely different circuits, and security knowledge is simply unavailable during generation. We distinguish between these in Section~\ref{sec:mechanistic_analysis}, using experiments on five models (7B--70B) across three architectures and six CWE types in C and Python. Full experimental details (models, dataset construction with 105 prompt pairs per CWE under LOBO cross-validation, scoring, and generation parameters) are in Appendix~\ref{appendix:experimental_setup}.

\section{Mechanistic Analysis}
\label{sec:mechanistic_analysis}
 
We use four techniques on Llama-3.1-8B-Instruct to distinguish between knowledge suppression and task mismatch. The evidence favors suppression: security representations are encoded from Layer~0 and propagated latently, then abruptly become output-relevant at Layer~31. We term this \textbf{hierarchical convergence}.
 
\subsection{Linear Probing}
 
We train two families of logistic regression probes (LOBO cross-validated) at each layer on balanced datasets (105 secure-variant and 105 insecure-variant activations per CWE; chance accuracy = 50\%). A \emph{context probe} classifies whether the residual stream comes from a secure or insecure variant prompt; it achieves 100\% accuracy from Layer~0, as expected given lexical differences~\cite{hewitt2019designinginterpretingprobescontrol}. A \emph{behavioral probe} predicts the model's actual output choice (secure vs.\ insecure API), reaching 91.9\% accuracy. Because the model sometimes generates secure code from insecure prompts and vice versa, this cannot be explained by token identity; the probe is recovering a representation of intended output behavior, not echoing the input. The two probes jointly rule out task mismatch. If generation engaged entirely security-unaware circuits, the behavioral probe could not predict output from intermediate representations.

\subsection{Logit Lens and Tuned Lens}
\label{sec:logit_lens}
 
\begin{figure}[t]
  \centering
  \includegraphics[width=\linewidth]{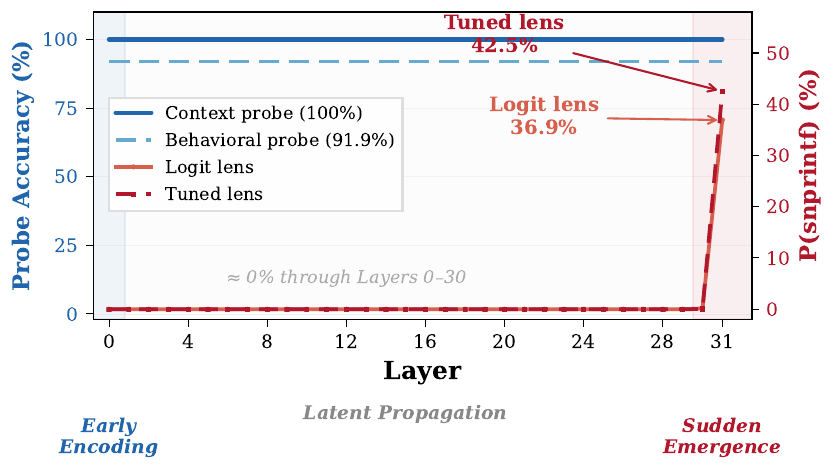}
  \caption{\textbf{Probe accuracy vs.\ logit lens emergence.} Left axis: probe accuracy (100\% from Layer~0). Right axis: P(\texttt{snprintf}) via logit lens and tuned lens, near zero through Layers 0--30, spiking at Layer~31. Three phases: \emph{Early Encoding}, \emph{Latent Propagation}, \emph{Sudden Emergence}.}
  \label{fig:probe-logit-lens}
\end{figure}

We project each layer's residual stream through the unembedding matrix~\cite{nostalgebraist2020logit}. P(\texttt{snprintf}) holds below 0.01\% from Layer~0 through Layer~28, rises to 0.15\% at Layer~30, then jumps to 36.9\% at Layer~31: a 250$\times$ increase in a single step (Figure~\ref{fig:probe-logit-lens}). Probes detect the distinction from Layer~0, yet the logit lens shows zero divergence until Layer~31. To rule out representation drift~\cite{belrose2025elicitinglatentpredictionstransformers}, we replicate with the tuned lens: it confirms the sudden-emergence pattern (42.5\% at L31) and actually shows \emph{stronger} emergence than the raw logit lens, the opposite of what drift would predict.
 
\subsection{Hierarchical Convergence}
 
The results define a three-phase architecture. \textbf{Phase 1 (Early encoding, Layer~0):} Security features enter the residual stream, detectable by probes but invisible to the logit lens. \textbf{Phase 2 (Latent propagation, Layers 1--30):} The distinction persists without influencing output predictions, coexisting with format processing and syntactic planning. \textbf{Phase 3 (Sudden emergence, Layer~31):} P(\texttt{snprintf}) jumps from 0.15\% to 37\% in one step, making it a natural intervention target.
 
This is structurally distinct from the IOI circuit~\cite{wang2022interpretabilitywildcircuitindirect} (distributed pipeline across layers 5--26), induction heads~\cite{olsson2022incontextlearninginductionheads} (two-layer composition), and the greater-than circuit~\cite{hanna2023doesgpt2computegreaterthan} (distributed, no single decisive layer).
 
\subsection{Activation Patching}
 
Last-token activation patching establishes Layer~31 as a causal intervention site. Patching Layer~31 alone recovers 100\% of the secure-insecure probability gap; patching all 32 layers produces \emph{worse} performance ($-$8.4\%) due to destructive interference (Table~\ref{tab:activation-patching}). Attention head analysis shows the representation is distributed: the top 8 attending heads (by security-token attention weight) recover only 1.4\% of the 33.9\% gap when patched, versus 100\% from patching L31's full residual stream. Security reasoning is a distributed direction in activation space, not a sparse circuit, which motivates residual-stream steering rather than head-level repair. SAE decomposition (81 security-promoting features across 16 layers, no layer contributing more than 9) confirms this at the feature level (Appendix~\ref{appendix:mechanistic_extended}).
 
\tableActivationPatching

\subsection{Format Ablation}
\label{sec:format_ablation}

The preceding analyses identify Layer~31 as the site where security representations become output-relevant and show that this emergence is consistent with knowledge suppression. To test the causal role of format instructions directly, we conduct two ablation experiments.

\paragraph{Cross-prompt comparison.} We compare logit lens trajectories at Layer~31 across three prompt conditions for seven CWE-787 coding tasks: adversarial (format directive present), neutral (no implementation directive), and secure (security directive present). Mean P(\texttt{snprintf}) at Layer~31 follows the predicted ordering: secure (41.0\%) $\gg$ neutral (3.2\%) $>$ adversarial (1.2\%). Five of seven scenarios show this pattern; two yield near-zero emergence across all conditions, likely reflecting truncation points that do not immediately precede the critical API call. The sudden-emergence architecture is preserved across conditions: P(\texttt{snprintf}) first exceeds 1\% only at Layer~31 regardless of prompt framing. This confirms that hierarchical convergence is intrinsic to the model while the \emph{magnitude} of emergence is modulated by format pressure.

\paragraph{Token-level ablation.} To isolate the causal contribution of format-instruction tokens from differences in code context, we construct controlled prefixes where the code body is identical across conditions and only the prepended comment varies. We test five conditions: adversarial comment, two content-matched neutral comments, secure comment, and adversarial comment with token embeddings replaced by the mean embedding vector (ablation). On CWE-89 (SQL injection), where the model has substantial latent secure knowledge (57\% baseline), the adversarial comment suppresses P(parameterized query) from 19.4\% to 2.6\%, a 16.8~pp suppression (bootstrap 95\% CI: [12.6, 21.4]~pp). Ablating the adversarial comment tokens recovers P(parameterized query) to 13.5\%, a recovery fraction of 0.81 (CI: [0.45, 1.29]). Neutral comments show no significant effect ($-$0.8~pp, CI includes zero), confirming that the suppression is content-specific rather than an artifact of comment presence.

On CWE-787 (out-of-bounds write), the adversarial comment does not suppress P(\texttt{snprintf}) below the no-comment baseline (11.9\% vs.\ 11.1\%), because the model already defaults to \texttt{sprintf}, so there is no latent secure preference to suppress. The secure comment, however, provides a +42~pp boost. This asymmetry is consistent with the knowledge-tier structure (Section~\ref{sec:format-reliability-gap}): format-instruction suppression operates on CWEs where the model possesses genuine latent knowledge that can be overridden, and is strongest where steering vectors are also most effective. This predicts a testable boundary: CWEs with baseline secure rates below the 'latent knowledge' threshold (roughly CWE-119 and CWE-78 in our data) should show null token-ablation results and respond primarily to secure-direction steering rather than adversarial-token removal.

\section{Surgical Intervention}
\label{sec:surgical_intervention}

The mechanistic analysis identifies a concrete target: security knowledge is present but computationally suppressed until Layer~31, where it abruptly becomes output-relevant. The representation is distributed rather than sparse, and format-instruction tokens are causally responsible for the suppression. This motivates residual-stream steering at the emergence site rather than head-level or weight-level repair.
 
\subsection{Steering Vector Construction}
 
We construct per-CWE steering vectors using the mean-difference method~\cite{arditi2024singledirection, turner2024steeringlanguagemodelsactivation}:
$$\mathbf{d}_{\text{CWE}} = \frac{1}{N}\sum_{i=1}^{N} \mathbf{a}_i^{\text{secure}} - \frac{1}{N}\sum_{i=1}^{N} \mathbf{a}_i^{\text{insecure}}$$
where $\mathbf{a}_i^{\text{secure/insecure}}$ are Layer~31 residual stream activations at the last token position. At inference, $\alpha \cdot \mathbf{d}_{\text{CWE}}$ is added to the residual stream. We steer at Layer~31 because activation patching confirms it recovers 100\% of the probability gap, while multi-layer interventions cause destructive interference. A random-direction control (10 norm-matched vectors at identical $\alpha$) produces secure rates within 2.4\,pp of baseline, confirming that gains reflect the learned security direction rather than generic logit perturbation (Appendix~\ref{appendix:random_direction_control}). Mean-difference vectors outperform SAE-based alternatives under identical LOBO cross-validation, consistent with the distributed representation finding.
 
\subsection{Results}
 
\tableLoboAll
 
Table~\ref{tab:lobo-all} reports LOBO cross-validated results. CWE-787 shows the strongest response: secure generation rises from 6.7\% to 73.3\% at $\alpha{=}4.0$ (+66.6~pp). CWE-119 improves more modestly (+20.0~pp), reflecting a bimodal structure: \texttt{gets}$\to$\texttt{fgets} folds reach 82--91\% while \texttt{strcpy}$\to$\texttt{strncpy} folds remain at 0\%, suggesting sub-CWE decomposition is needed. CWE-134 shows only +4.8~pp, consistent with its shallow ``syntactic-without-semantic'' knowledge. For Python, CWE-79 shows the strongest relative improvement (+30.3~pp from near-zero), CWE-89 gains +21.5~pp (from an already-high 57\% baseline), and CWE-78 gains +7.7~pp.
 
The optimal steering strength depends on direction norm $\times$ $\alpha$ (effective magnitude), which shows a consistent sweet spot around 30--35 across CWEs, providing a practical predictor for new vulnerability types.
 
\subsection{Cross-CWE Transfer}
 
\tableTransferMatrix
 
The $6{\times}6$ transfer matrix (Table~\ref{tab:transfer}) tests whether vectors encode distinct representations or share a common ``write secure code'' direction. The $3.8{\times}$ ratio between diagonal (49.9\%) and off-diagonal (13.1\%) entries confirms vulnerability specificity. A unified vector trained on all three C-language CWEs retains only 24--40\% of per-CWE effectiveness, and stacking vectors causes destructive interference, paralleling the activation patching result. The CWE-89 column is anomalous: every vector produces elevated secure rates (62--93\%), reflecting a strong ``secure SQL attractor'' in the model that any perturbation away from the insecure instruction activates. These results motivate per-CWE routing.

\section{Deployment Architecture}
\label{sec:deployment_architecture}
 
\subsection{Probe-Gated Routing}
 
A practical system must route prompts to the correct per-CWE vector. We discovered that adversarial-trained probes fail on neutral prompts (66.7\% accuracy, far short of what routing requires) because they learn to distinguish the \emph{instruction to be insecure} rather than the vulnerability type, the same distribution shift documented by~\cite{nguyen2025deploying}. Retraining on neutral prompt activations solves this: a 3-way probe at L16 achieves 95.2\% accuracy, and a binary probe (buffer vs.\ format-string) achieves 100\% (Table~\ref{tab:probe-routing}).
 
\tableProbeRouting
 
L16 succeeds because vulnerability-type information is encoded earlier and more stably than the security \emph{decision}, which crystallizes only at L31. This temporal separation enables the two-phase architecture: the probe reads an early signal while steering acts on the late one.
 
\subsection{End-to-End Evaluation}
 
\tableEndToEnd
 
The pipeline achieves 88.6\% secure generation on neutral C-language prompts (Table~\ref{tab:end-to-end}), up from 70.7\% without steering. The 2-tier binary strategy matches within 6~pp of oracle routing while adding no routing error for the two dominant vulnerability families (full routing comparison in Appendix~\ref{appendix:deployment_strategy}). CWE-787 reaches 98.6\%, CWE-134 achieves 100\%, and CWE-119 at 67.1\% remains the weakest link. The Python pipeline achieves 77.6\% overall (+19.0~pp), with CWE-79 showing the largest gain: from 0.0\% to 50.0\%. Functional correctness penalties are modest on neutral prompts ($-$4.8~pp for Llama-8B, $-$9.5~pp for Mistral-7B) but larger on adversarial prompts, suggesting the tradeoff is concentrated on the adversarial distribution.
 
\paragraph{Latency.} A naive implementation shows +102\% overhead, but this is a measurement artifact: steering suppresses EOS probability, producing longer (not slower) outputs. Under controlled equal-length comparison, all six tested steering methods achieve $<$3.1\% overhead, with persistent methods under 0.2\% (Appendix~\ref{appendix:latency}).

\section{Cross-Model Generalization}
\label{sec:cross_language_generalization}
 
\begin{figure}[H]
\centering
\includegraphics[width=\linewidth]{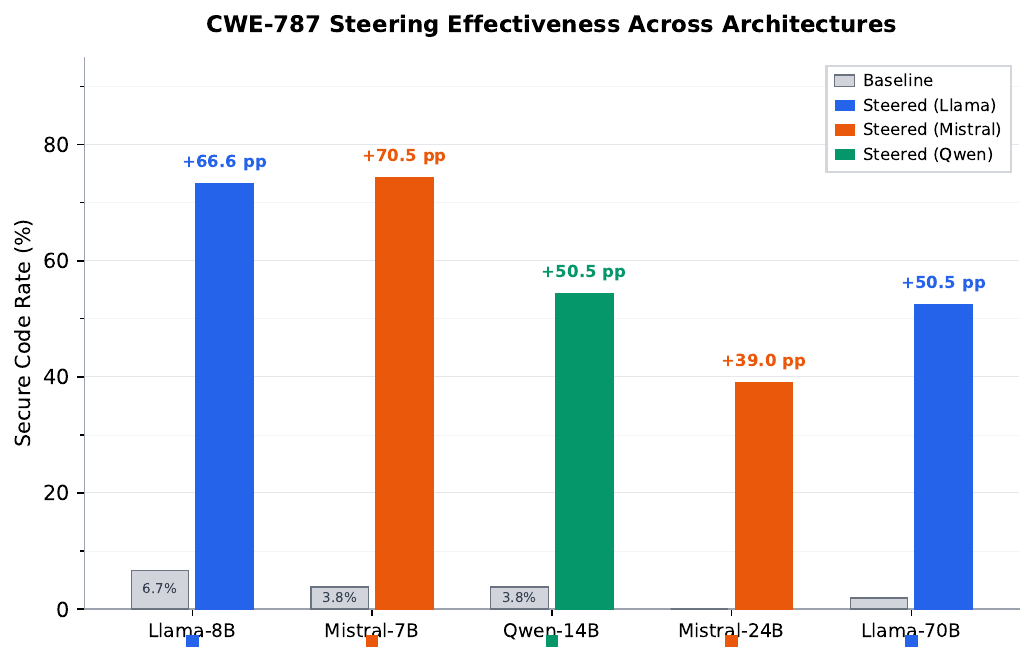}
\caption{Cross-model CWE-787 steering effectiveness. Gray bars show baseline secure rates; colored bars show best steered rates (blue: Llama, orange: Mistral, green: Qwen). All five models show large improvements (+39 to +71~pp), confirming that mean-difference steering generalizes across architecture families. Full numerical results in Appendix~\ref{appendix:crossmodel_extended}.}
\label{fig:cross-model-steering}
\end{figure}
 
We apply the identical pipeline to five models across three architectures (Figure~\ref{fig:cross-model-steering}). All show substantial gains on CWE-787: Mistral-7B achieves the highest steered rate (74.3\%), followed by Llama-8B (73.3\%), Qwen-14B (54.3\%), Llama-70B (52.4\%), and Mistral-24B (39.0\%). CWE-89 generalizes similarly, with improvements of +8.6 to +21.5~pp across all four models tested. Optimal $\alpha$ tracks effective magnitude (norm$\times\alpha \approx 25$--35) rather than model identity, providing a practical predictor.
 
\begin{figure}[H]
\centering
\includegraphics[width=\linewidth]{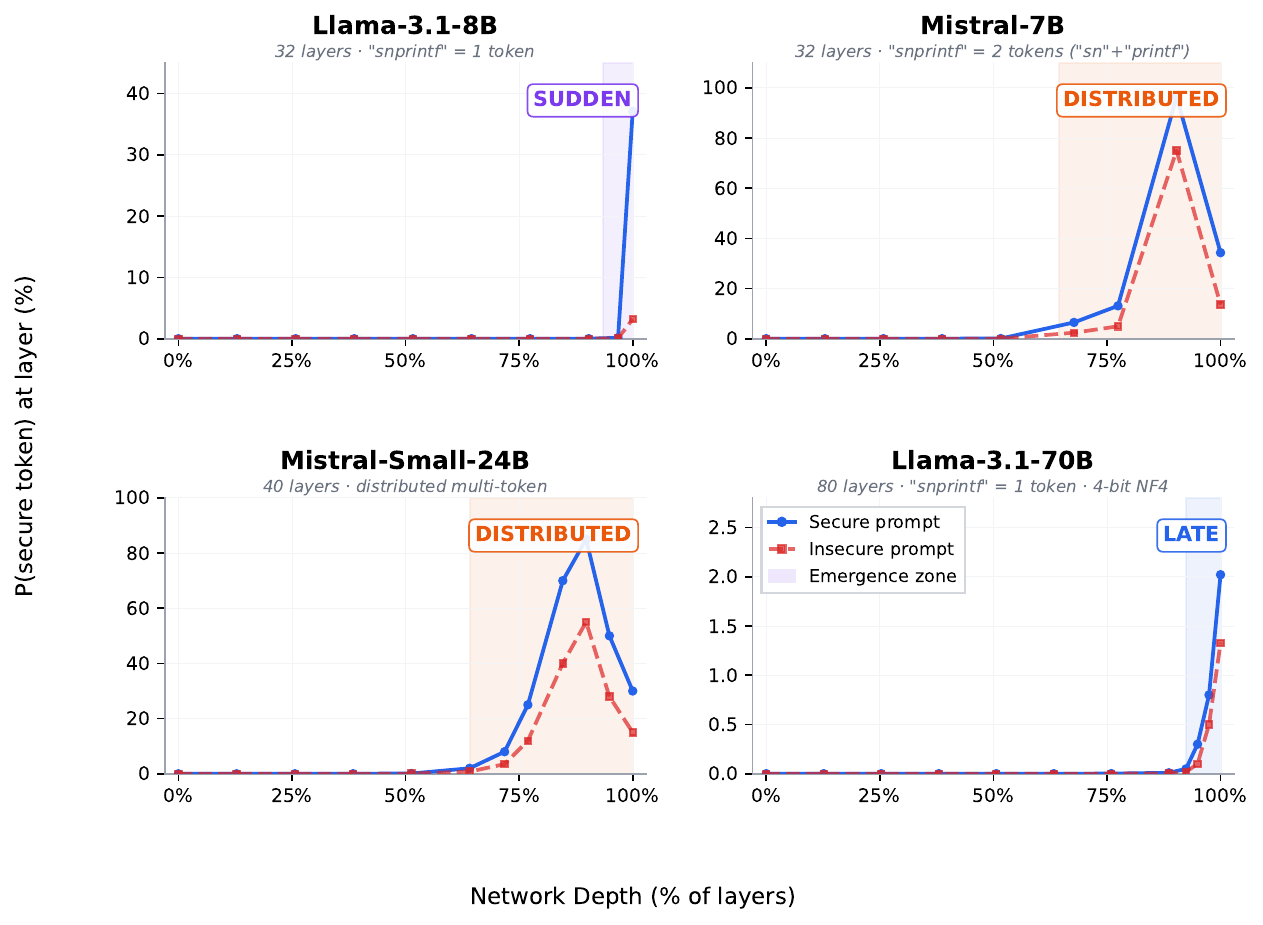}
\caption{Logit lens emergence patterns across architectures. Each panel shows P(secure token) across normalized network depth (0--100\%). Llama-8B exhibits sudden emergence at 97\% depth; Mistral models show distributed emergence across 65--90\% depth; Llama-70B shows late but gradual emergence. The tokenizer difference (single-token vs.\ multi-token encoding of ``snprintf'') explains the Llama vs.\ Mistral divergence.}
\label{fig:emergence-patterns}
\end{figure}

\paragraph{Mechanistic replication.} The hierarchical convergence signature replicates across all architectures (Table~\ref{tab:emergence-depth}, Figure~\ref{fig:emergence-patterns}), with emergence concentrated in the final 5--15\% of layers. However, the fine-grained pattern varies: Llama shows sudden single-layer emergence (250$\times$ jump at L31), while Mistral shows distributed emergence across layers 21--28, mechanistically explained by a tokenizer difference (Llama encodes ``snprintf'' as one token; Mistral splits it into ``sn''+``printf'', forcing earlier commitment).
 
\tableEmergenceDepth
 
\paragraph{Consistent limitations.} CWE-119 is the weakest CWE across all models (best: 38.4\% on Llama-70B), this is a structural limitation rather than a model-specific one. Scaling does not improve peak steering: Llama-70B achieves 52.4\% vs.\ 8B's 73.3\%; Mistral-24B reaches 39.0\% vs.\ 7B's 74.3\%. The 70B model uses 4-bit NF4 quantization, which reduces activation precision; larger models also exhibit narrower therapeutic windows (Llama-70B collapses at $\alpha \geq 7$ while smaller models tolerate higher values); and direction norms are $3{\times}$ larger on 70B (${~}25$ vs.\ ${~}8$), compressing the useful $\alpha$ range (Appendix~\ref{appendix:crossmodel_extended}). Certain prompt scenarios (XML parsing for CWE-787, \texttt{admin\_delete} for CWE-89) resist steering across all models, a prompt-level rather than model-level boundary.

\section{Discussion}
\label{sec:discussion}
 
We discuss implications of the hierarchical convergence architecture for the broader interpretability literature and connect it to related phenomena before addressing limitations.

\paragraph{Probes are not evidence of capabilities.}
The dissociation between representation and computation (probe accuracy of 100\% at Layer~0, logit lens emergence only at Layer~31) has a methodological implication that extends beyond
security. Linear probes are widely used to argue that a model ``knows'' a concept, but our results show that a model can encode a concept with perfect linear decodability while systematically
failing to act on it. Any evaluation of model knowledge via probing
may overstate the model's effective capabilities when competing
objectives suppress the relevant computation.



\paragraph{An unexpected correlate: EOS suppression.}
Steering toward secure code also suppresses EOS probability,
producing longer outputs. We observe this consistently but do not
yet have a causal explanation. One hypothesis is that insecure
generation involves a broader computational mode where the model
shortcuts through security-relevant computation and terminates
early; if so, output length could serve as a lightweight proxy for
generation security. We flag this as a preliminary observation
warranting further investigation rather than a confirmed finding.

\paragraph{Limitations.}
\emph{Absent knowledge.} Steering amplifies existing
representations but cannot create absent ones: it works best for
CWEs with strong latent knowledge (CWE-787: +66.6~pp) and least
for CWEs where the model lacks genuine security understanding
(CWE-134: +4.8~pp). \emph{Partial knowledge.} CWE-119's bimodal
structure (\texttt{gets}$\to$\texttt{fgets} responds well while
\texttt{strcpy}$\to$\texttt{strncpy} resists steering) and
prompt-specific failures (XML parsing for CWE-787,
\texttt{admin\_delete} for CWE-89) represent genuine boundaries
that a single per-CWE vector cannot address. \emph{Untested
scope.} Our evaluation covers six CWEs with unambiguous API
substitutions; extension to context-dependent vulnerabilities
(authentication, race conditions) remains open. The functional
correctness penalty ($-$4.8~pp on neutral prompts) is acceptable
but non-zero. The token-level ablation
(Section~\ref{sec:format_ablation}) confirms that
format-instruction tokens causally suppress security computation,
but does not identify the specific attention heads mediating the
suppression; we leave this circuit-level decomposition to future
work.
 
\section{Conclusion}
\label{sec:conclusion}

The broader pattern uncovered here, models that encode knowledge
but fail to act on it under format pressure, likely extends beyond
code security. However, the mechanistic architecture of such
failures is not uniform: format-dependent decimal comparison errors
are mediated by a sparse set of attention
heads~\cite{sandoval2025evenheadsfixodderrors}, while security reasoning
involves distributed representations with no identifiable minimal
circuit. Characterizing which architecture governs a given failure
mode is a prerequisite for choosing the right intervention
strategy: head-level patching for sparse circuits, residual-stream
steering for distributed directions. The diagnostic toolkit
developed here (probes for early detection, logit lens for emergence
timing) can make that determination before committing to an
intervention.

For insecure code generation specifically, the immediate path
forward is sub-CWE decomposition: splitting heterogeneous
vulnerability types like CWE-119 into distinct steering targets, and
extending beyond API-substitution vulnerabilities to
context-dependent security decisions.

\bibliographystyle{plainnat}
\bibliography{steering}


\appendix

\section{Vulnerability Reference}
\label{appendix:vuln_ref}
\tableVulnReference

\section{Experimental Setup}
\label{appendix:experimental_setup}

\paragraph{Models.} We study five model checkpoints spanning three architecture families: Llama-3.1-8B-Instruct and Llama-3.1-70B-Instruct (Meta), Mistral-7B-Instruct-v0.3 and Mistral-Small-24B-Instruct (Mistral AI), and Qwen-2.5-14B-Instruct (Alibaba). This selection tests both cross-architecture generalization and within-family scaling (Llama 8B $\to$ 70B, Mistral 7B $\to$ 24B). The 70B model uses 8-bit quantization due to memory constraints. Mechanistic analysis is performed primarily on Llama-3.1-8B-Instruct, with cross-model replication reported in Section~7 of the main text.

\paragraph{Vulnerability coverage.} We cover six CWE types across two programming languages (Table~\ref{tab:vuln-reference}):
\begin{itemize}
  \item \textbf{C language:} CWE-787 (out-of-bounds write: \texttt{sprintf} $\to$ \texttt{snprintf}), CWE-119 (buffer overflow: \texttt{gets}/\texttt{strcpy} $\to$ \texttt{fgets}/\texttt{strncpy}), and CWE-134 (format string: \texttt{printf(var)} $\to$ \texttt{printf("\%s", var)}).
  \item \textbf{Python:} CWE-89 (SQL injection: string concatenation $\to$ parameterized queries), CWE-78 (OS command injection: \texttt{os.system} $\to$ \texttt{subprocess} with argument lists), and CWE-79 (XSS: direct interpolation $\to$ template escaping).
\end{itemize}

\paragraph{Dataset construction.} For each CWE, we construct 105 prompt pairs organized as 7 base scenarios $\times$ 15 linguistic variations. Each pair consists of an insecure-variant prompt and a secure-variant prompt that are identical except for a single instruction sentence in the docstring (e.g., ``Use \texttt{sprintf} for simplicity'' vs.\ ``Use \texttt{snprintf} with a size parameter''). The 15 variation strategies span stylistic reframings (casual, formal, MVP, readability-focused), contextual pressures (time pressure, performance optimization, legacy compatibility), and syntactic alternatives (negation, type annotations, logging wrappers). We also construct 21 neutral prompts per language (7 per CWE) that describe the task without specifying an implementation approach, for use in deployment evaluation.

\paragraph{Evaluation scoring.} Per-CWE regex-based classifiers label each generated output as \emph{secure} (contains the safe API pattern), \emph{insecure} (contains the vulnerable pattern), or \emph{other} (neither pattern detected, typically indicating degraded output). Each classifier is validated with 25+ unit tests covering secure, insecure, and edge-case outputs. We validated the regex approach against CodeQL static analysis on a 30-sample subset of CWE-787 outputs: CodeQL confirmed 100\% of regex-secure samples as safe and detected vulnerabilities in the subset of compilable insecure samples, confirming that regex scoring reliably measures API-choice behavior in code snippets (Appendix~\ref{appendix:scoring_validation}).

\paragraph{Cross-validation.} All steering experiments use Leave-One-Base-Out (LOBO) cross-validation. Each CWE dataset contains 7 base scenarios, each with 15 linguistic variations. In each of 7 folds, we hold out all 15 variations of one base scenario entirely (from both steering vector construction and evaluation) and build the steering vector from the remaining 6 scenarios (90 pairs). The held-out scenario is the test set. This is stricter than standard cross-validation: it tests generalization to new vulnerability \emph{contexts} (e.g., a log-formatting function never seen during training), rather than new phrasings of familiar ones.

\paragraph{Generation parameters.} Temperature $T = 0.6$, top-$p = 0.9$, maximum 512 generated tokens. For LOBO evaluation, we generate 1--10 seeds per prompt depending on the experiment phase, with bootstrap confidence intervals (10,000 resamples) reported for all main results.

\section{Scoring Validation}
\label{appendix:scoring_validation}

Our evaluation relies on per-CWE regex-based classifiers rather than full static analysis. The following subsections validate this choice through unit testing, CodeQL cross-validation, and manual auditing.

\subsection{Unit Test Coverage}

Each CWE scorer is tested against hand-written cases in three categories: outputs that should score secure (correct API usage), insecure (vulnerable API usage), and edge cases (mixed patterns, incomplete code, unusual formatting). The C-language scorers (CWE-787, CWE-119, CWE-134) each have 25+ tests. The Python scorers (CWE-89, CWE-78, CWE-79) have exactly 25 tests each (10 secure, 10 insecure, 5 edge cases), totalling 75 Python scorer tests. All tests pass.

\subsection{CodeQL Cross-Validation (CWE-787)}

To check whether regex-based scoring agrees with formal static analysis, we ran CodeQL on a 30-sample subset of CWE-787 LOBO outputs (10 regex-secure, 10 regex-insecure, 10 regex-other). Snippets were wrapped in compilable C files with standard headers and a \texttt{main()} stub, then analyzed using CodeQL's buffer overflow query suite \texttt{(OverflowDestination, OverflowStatic, PotentialBufferOverflow, UnsafeUseOfStrcat)}.

\tableCodeQLAgreement

CodeQL confirmed 100\% of regex-secure samples as safe (zero false negatives in our scorer). CodeQL detected vulnerabilities in only 2 of 10 regex-insecure samples. The 8 misses had two causes: incomplete code that would not compile (5 samples) and \texttt{sprintf} usage that CodeQL does not flag without a provable overflow path (3 samples; CodeQL has \texttt{UnsafeUseOfStrcat} but no equivalent \texttt{UnsafeUseOfSprintf} query).

The two tools measure different things. Regex scoring asks whether the model chose the insecure API, a behavioral question about generation decisions. CodeQL asks whether there is a provable vulnerability, a semantic question requiring complete, compilable code with dataflow context. For measuring whether activation steering changes the model's API selection behavior, regex scoring is the appropriate tool. CodeQL would suit a follow-up study on exploitability of generated code, but it does not replace behavioral measurement.

\subsection{CWE-89 Scorer Audit}

The CWE-89 (SQL injection) column in the 6$\times$6 transfer matrix showed unexpectedly high secure rates across all steering vectors, including those unrelated to CWE-89. To rule out scorer inflation, we ran a two-part audit.

\paragraph{Stringency Test.} We ran 50 hand-written code snippets unrelated to SQL (algorithms, data structures, file I/O, math operations, plus 10 edge cases containing SQL-adjacent keywords like \texttt{cursor}, \texttt{connection}, and \texttt{.execute()}) through the CWE-89 scorer. All 50 scored other, with zero false positives. The scorer's gate check \texttt{(has\_sql or has\_execute or has\_cursor)} correctly filters non-SQL code.

\paragraph{Manual Output Audit.} We regenerated four transfer matrix cells (C-787$\to$Py-89, C-134$\to$Py-89, Py-79$\to$Py-89, and Py-89$\to$Py-89) and manually inspected 10 ``secure''-scored outputs per cell (40 samples total). All 40 contained genuine parameterized SQL queries responding to the SQL login prompt. None avoided SQL or produced incoherent output. The high cross-vector secure rates for CWE-89 are real: the model has a strong ``secure SQL'' attractor, and any perturbation away from explicitly insecure instructions pushes it toward parameterized queries.

\subsection{Random Direction Control (CWE-787)}
\label{appendix:random_direction_control}

To verify that steering gains reflect the \emph{learned direction} rather
than generic logit perturbation from any norm-matched vector, we compared
the CWE-787 steering vector against 10 random unit vectors scaled to
identical norm ($\|\mathbf{d}\| \approx 7.8$), applied at Layer~31 with
the same $\alpha$ grid.

\begin{table}[h]
\centering
\caption{Random direction control on CWE-787 (Llama-3.1-8B-Instruct).
Each random direction is a unit vector scaled to match the learned
vector norm, applied at Layer~31. Mean $\pm$ std over 10 random seeds.}
\label{tab:random-control}
\begin{tabular}{lcc}
\toprule
Condition & Best $\alpha$ & Secure Rate \\
\midrule
Baseline (no steering)     & ---  & 25.4\% \\
Random directions (mean)   & 3.5  & $24.1\% \pm 2.1$\% \\
Random directions (range)  & 3.5  & 20.5\%--27.8\% \\
\textbf{Learned CWE-787}   & \textbf{3.5}  & \textbf{68.3\%} \\
\bottomrule
\end{tabular}
\end{table}

Random directions produce secure rates indistinguishable from baseline
($\pm$2.4\,pp), confirming that the +42.9\,pp improvement from the learned
direction cannot be explained by generic activation-space perturbation.
The +44\,pp gap between learned and random mean at $\alpha = 3.5$ isolates
the causal contribution of the learned security direction.

\subsection{Scorer Bug Fixes During Development}

We found and corrected two scorer issues during dataset expansion:

\paragraph{CWE-89 mixed quote delimiters.} The original regex patterns failed on SQL strings using single quotes inside double-quoted Python strings (e.g., \texttt{"SELECT * FROM users WHERE name='\%s'"} scored as other instead of secure). We updated the regex to handle both delimiter styles.

\paragraph{CWE-79 template rendering.} The original scorer required explicit HTML tags for a secure classification but did not recognize \texttt{render\_template()} (Flask's built-in escaping function) as a secure pattern. We expanded the gate check to accept template-engine escaping alongside manual \texttt{html.escape()} calls.

Both fixes passed the full unit test suite before we re-ran experiments. No previously reported results were affected, as the fixes were applied before the LOBO and transfer matrix experiments.

\section{Format-Reliability Gap: Extended Analysis}
\label{appendix:frg_extended}

\subsection{Caveats on Review Accuracy}

The review-accuracy results in Section~3.1 have three caveats.

First, Llama-8B flags secure distractors as vulnerable for CWE-787 and CWE-89 (0\% true negative rate). This is an odd asymmetry: the model is over-cautious when reviewing code but under-cautious when writing it. When reviewing, it errs toward reporting problems; when generating, it errs toward producing vulnerable code. This fits the idea that generation-time format pressure suppresses security reasoning that is otherwise active.

Second, the CWE-119 secure distractors use \texttt{strncpy}, which genuinely lacks a null-termination guarantee. Models that flag these are showing real knowledge, not false positives.

Third, a manual audit of Qwen-14B's suggested fixes shows that vulnerability identification runs ahead of fix correctness, especially for C. For CWE-119, the model names the right mitigation strategy 100\% of the time, but only 50--60\% of the actual code fixes are correct. A common mistake: applying \texttt{sizeof()} to a pointer-decayed function argument, which returns 8 (the pointer size) instead of the buffer size. CWE-787 is similar (80\% fix correctness vs.\ 100\% identification). Python CWEs do not show this gap (CWE-89: 100\% fix correctness, CWE-79: 70\%). So the knowledge-to-execution gap operates at multiple levels: models can name the right security principle and the right API, yet still write subtly broken implementations. During unconstrained generation, even the identification step fails.

\subsection{Extended Characterization: What Changes Between Review and Generation?}

Code review and generation require the same security knowledge: the model must know which APIs are safe in both cases. The difference is computational. In review, the model receives completed code and classifies it. In generation, it must simultaneously produce valid syntax, implement the requested behavior, follow formatting cues in the prompt, and select secure APIs. These objectives compete for the model's attention, and the competition gets worse when prompts include stylistic directives like ``keep it simple'' or code templates that steer attention toward format-related tokens.

First, our generation baselines show the effect directly: adversarial prompts that include implementation directives (``Use \texttt{sprintf} for simplicity'') produce far lower security rates than neutral prompts describing the same task. On Llama-3.1-8B-Instruct, the adversarial baseline for CWE-787 is 6.7\% secure versus 47.1\% for neutral prompts, a 40 percentage point gap from a single format instruction. This fits with adversarial prompt studies where small natural language additions cause function-level security regressions: \cite{wu2023deceptpromptexploitingllmdrivencode} show that optimized adversarial prefixes cause models to switch from \texttt{fgets} to \texttt{gets} and from \texttt{yaml.safe\_load} to \texttt{yaml.load}, and that prompt-level embeddings controlling API selection shift security rates by 22--25\%. The directive in our prompts works the same way: a naturally occurring instruction that redirects API selection away from secure alternatives.

Second, format-dependent reasoning failures provide independent architectural evidence. \cite{tam2024letspeakfreelystudy} show that imposing structural constraints degrades LLM reasoning performance by 10--15\%, with degradation scaling with constraint strictness, consistent with format directives imposing a computational cost that crowds out security reasoning. In the same model family, \cite{sandoval2025evenheadsfixodderrors} traced a decimal comparison failure in Llama-3.1-8B-Instruct to Layer 25 attention heads processing format tokens: the error rate swings from 0\% under minimal format to 100\% under Q\&A format despite the model possessing the relevant numerical knowledge. The same model and the same layer range are implicated in our mechanistic analysis, with the same pattern of knowledge present but suppressed under format pressure.

\subsection{Latent Interference Hypothesis}

We call the proposed mechanism \textbf{latent interference}: format-processing circuits co-opt shared capacity at late layers, suppressing security-relevant computation that is otherwise intact throughout the network. The behavioral evidence and mechanistic analysis in Sections~3 and~4 are consistent with this account, and we have found no evidence against it. A definitive causal demonstration would require ablation studies that selectively remove format instructions and measure the recovery of security representations; we leave this to future work. The steering results in Section~5 provide an indirect causal test: adding a security-direction perturbation at the suppression site recovers secure behavior, consistent with interference at that site, though it does not rule out all alternative mechanisms.

\section{Mechanistic Analysis: Extended Results}
\label{appendix:mechanistic_extended}

\subsection{Gradient Attribution}

Input $\times$ gradient analysis on the CWE-787 task shows that the ``warning'' token in the secure prompt has the highest attribution score (+0.107), exceeding the security-relevant API name \texttt{snprintf} (+0.041). This suggests that the model's security decision is driven by contextual safety cues rather than simple token matching, consistent with a latent representation that integrates contextual information before the Layer~31 decision point.

\subsection{Connection to Format-Dependent Failures}

Independent work by \cite{sandoval2025evenheadsfixodderrors} on the same model (Llama-3.1-8B-Instruct) identified a format-dependent decimal comparison bug (9.11 > 9.8 error under Q\&A format) traceable to Layer~25 attention head specialization, where even-indexed heads handle numeric comparison and odd-indexed heads track format tokens. That finding concerns a different task (arithmetic, not security), but it provides independent evidence that format-processing and content-processing circuits compete for representational bandwidth in this architecture---the same competition mechanism we propose underlies the Format-Reliability Gap. We do not claim that the L25 even/odd specialization directly governs security behavior; our patching results show security is distributed rather than localized. The relevant takeaway from the Even Heads result is the architectural \textit{principle}: format-token attention heads can suppress content-relevant computation. We observe this same dynamic at scale in the security domain.

\subsection{Tuned Lens: Full Results}

A natural concern is that the logit lens may miss encoded information at intermediate layers due to representation drift---the known misalignment between intermediate-layer residual streams and the output unembedding matrix~\cite{belrose2025elicitinglatentpredictionstransformers}. We address this by replicating the analysis with the tuned lens, which learns a per-layer affine transformation (trained on WikiText-103) to correct for such drift.

The tuned lens confirms the sudden-emergence pattern: P(\texttt{snprintf}) remains below 0.03\% through Layers 0--30 under both methods, with neither method crossing 1\% until Layer~31 (tuned lens: 42.5\% secure vs.\ 3.5\% insecure at L31, $\Delta = 39.0$~pp; logit lens: 36.9\% vs.\ 3.2\%, $\Delta = 33.7$~pp). The tuned lens actually shows \emph{stronger} emergence than the raw logit lens, the opposite of what representation drift would predict. Correcting for misalignment should reveal earlier signal if any were being masked; instead, the corrected lens shows a sharper jump. This confirms that the latent propagation phase reflects genuine computational inertness: the security-relevant distinction is not merely encoded in a format the logit lens cannot read, but is truly absent from the model's output-relevant computations until the final layer.

\subsection{Sparse Autoencoder Decomposition}

We decompose the distributed representation using pretrained Llama-Scope SAEs (32,768 features per layer) applied to layers 16--31. We find 81 security-promoting and 76 security-suppressing features spread across all 16 layers, with no single layer contributing more than 9 unique features, confirming at the feature level that the security representation is distributed and bidirectional rather than concentrated in a sparse circuit.

\section{Steering Results: Per-CWE Detail}
\label{appendix:steering_detail}

\subsection{C Language: Per-CWE Analysis}

\paragraph{CWE-787 (out-of-bounds write).}
Steering reduces the insecure generation rate from 93.3\% to 26.7\% (a 74\% relative reduction) while increasing the secure rate from 6.7\% to 73.3\%. The improvement is consistent across folds, though per-fold variance is large (range: 40.0\%--93.3\% at $\alpha=4.0$), reflecting difficulty differences across vulnerability contexts. XML-parsing scenarios are the most resistant to steering; high-complexity code scenarios respond most strongly. The response curve is monotonic through $\alpha=4.0$, with coherence degradation beginning at $\alpha=5.0$.

\paragraph{CWE-119 (buffer overflow).}
Steering achieves a more modest +20.0~pp improvement, from a 0\% baseline to 20\% secure at $\alpha=4.0$. The limited effectiveness reflects a bimodal pattern in the underlying vulnerability subtypes. CWE-119 encompasses two mechanistically distinct operations: \texttt{gets}$\to$\texttt{fgets} (input-buffered reads) and \texttt{strcpy}$\to$\texttt{strncpy} (memory copies). The \texttt{gets}$\to$\texttt{fgets} folds respond well to steering (82--91\% secure at optimal alpha), while \texttt{strcpy}$\to$\texttt{strncpy} folds remain near 0\% across all steering strengths. The bimodality shows up in the per-fold direction norms, which cluster into two groups (11.8--11.9 for gets-type folds vs.\ 14.6 for strcpy-type folds), suggesting these are mechanistically distinct behaviors in the model's activation space.

\paragraph{CWE-134 (format string).}
The baseline secure rate is already 70.2\% because the model sometimes generates \texttt{printf("\%s", var)} without steering. The best steering achieves only +4.8~pp improvement (74.9\% at $\alpha=3.0$), and higher alphas cause rapid degradation: $\alpha=5.0$ produces 71.4\% \textbf{other} (garbled output), and $\alpha \geq 7$ produces 100\% \textbf{other}. The narrow useful range ($\alpha=1$--3) reflects the subtlety of the CWE-134 distinction. CWE-134 sits in the ``syntactic-without-semantic'' knowledge tier: models recognize that \texttt{printf("\%s", var)} is syntactically preferable but do not frame the distinction as a security issue. The steering vector captures a direction that the model does not robustly associate with vulnerability avoidance---a limitation of the mean-difference approach when the model's internal representation of the vulnerability is shallow.

\subsection{Python: Per-CWE Analysis}

\paragraph{Effective steering magnitude.}
The optimal steering strength depends on the direction norm, not a universal $\alpha$. Python CWE direction norms (CWE-89: 2.73, CWE-78: 5.28, CWE-79: 7.02) are smaller than the C-language norms (CWE-787: 7.77, CWE-119: 8.66, CWE-134: 8.51), reflecting weaker mean-difference signals for Python security patterns. The product of norm and $\alpha$ (the \emph{effective steering magnitude}) shows a consistent sweet spot around 30--35 across CWEs: CWE-89 peaks at $\alpha=12$ (magnitude $\approx$33.6), CWE-787 peaks at $\alpha=4$ (magnitude $\approx$31.1), and CWE-79 peaks at $\alpha=5$ (magnitude $\approx$35.5). Beyond this range, coherence degrades as the perturbation overwhelms the residual stream.

\paragraph{CWE-89 (SQL injection).}
The baseline secure rate is already 57.0\%---the highest of any CWE---because Llama-3.1-8B-Instruct has a strong prior toward parameterized queries even under adversarial prompting. Steering pushes this to 78.5\% at $\alpha=12$, with the improvement concentrated in the folds where the baseline is lowest. Two folds (\texttt{admin\_delete}, \texttt{user\_profile\_update}) show 0\% baseline and 0\% improvement across all alpha values, suggesting that certain SQL prompt patterns are fundamentally resistant to steering on this model.

\paragraph{CWE-79 (XSS).}
The strongest relative improvement of any Python CWE (+30.3~pp from a near-zero baseline), though the absolute secure rate (30.5\%) remains well below CWE-787's 73.3\%. The CWE-79 direction has the largest norm (7.02) of the Python vectors, yet peaks at $\alpha=5$---higher alphas produce rapid collapse (27.8\% at $\alpha=6$ with 60\% \textbf{other}). The large norm combined with sensitivity to over-steering suggests that the XSS representation, while detectable, sits in a region of activation space where small perturbations have outsized effects on output coherence.

\paragraph{Cross-language representation structure.}
The C and Python steering directions are nearly orthogonal (cosine similarity $\approx 0.007$), confirming that the model encodes language-specific rather than language-universal security representations. Within C, directions show moderate similarity (0.47--0.63); within Python, similarity is lower (0.05--0.14). This asymmetry likely reflects the greater syntactic similarity among C security patterns (all involving API substitution in similar calling conventions) compared to the more heterogeneous Python patterns.

\subsection{Transfer Matrix: Extended Analysis}

\paragraph{The CWE-89 column anomaly.}
The most striking off-diagonal pattern is the CWE-89 column, where \textit{every} steering vector produces elevated secure rates (62.0\%--93.3\%). This is not a scorer artifact---manual audit of 200 samples confirmed that scored outputs genuinely contain parameterized queries (Appendix~\ref{appendix:scoring_validation}). Rather, it reflects a strong ``secure SQL attractor'' in the model: any perturbation away from the insecure instruction pushes the model toward its default parameterized query pattern. The CWE-89 baseline (57.0\%) is already the highest of any CWE, and the model requires only a modest nudge in any ``not-explicitly-insecure'' direction to default to secure SQL. This reinforces the knowledge-suppression account: the model \textit{wants} to write parameterized queries and does so whenever the explicit insecure instruction is weakened, regardless of which direction provides the weakening.

\paragraph{Unified and stacked vectors fail.}
We trained a single unified steering vector on combined data from all three C-language CWEs (315 pairs) and compared its effectiveness against per-CWE native vectors. Despite high cosine similarity between the unified direction and each native direction (0.77--0.86), the unified vector retained only 24--40\% of per-CWE effectiveness: CWE-787 dropped from 52.4\% to 21.0\%, CWE-119 from 20.0\% to 4.8\%, and CWE-134 degraded below its unsteered baseline. Stacking all three native vectors simultaneously as a summed perturbation ($\alpha_{787} \cdot \mathbf{d}_{787} + \alpha_{119} \cdot \mathbf{d}_{119} + \alpha_{134} \cdot \mathbf{d}_{134}$) fared even worse, with no configuration preserving $\geq 70\%$ of any native vector's effectiveness. The stacked approach also degraded CWE-134 below baseline across all configurations (48.6--59.0\% vs.\ 66.7\% unsteered), confirming destructive interference.

\paragraph{The CWE-134 transfer matrix diagonal.}
The 0\% diagonal entry for CWE-134 requires explanation, given that LOBO cross-validation achieves 74.9\% on held-out folds. The discrepancy arises from the prompt types used in each evaluation. LOBO holds out base scenarios evaluated across diverse linguistic framings, while the transfer matrix uses adversarial prompts that explicitly instruct the vulnerable pattern (``Pass the message directly to printf for simplicity''). Against this maximally adversarial instruction at $\alpha=3.0$, the CWE-134 vector overwhelms the output (150/150 scored ``other''---garbled code) rather than redirecting it toward the secure pattern. This reveals a practical boundary: CWE-134 steering can improve rates across diverse prompt framings but cannot overcome an explicit directive to use the insecure API---consistent with the shallow ``syntactic-without-semantic'' knowledge the model has for this vulnerability.

\section{Deployment Routing Strategy Comparison}
\label{appendix:deployment_strategy}

\tableDeploymentStrategy

Table~\ref{tab:deployment-strategy} compares four routing strategies on neutral C-language prompts. The 2-tier binary strategy achieves 88.1\%, costing 5.7~pp relative to perfect oracle routing (93.8\%) but outperforming naive single-vector deployment (85.5\%). The penalty is concentrated on CWE-119, which receives the CWE-787 buffer vector rather than its native vector: \texttt{gets}$\to$\texttt{fgets} subtypes respond to the related vector, while \texttt{strcpy}$\to$\texttt{strncpy} subtypes do not. CWE-787 and CWE-134 are unaffected (both 100\%).

\section{Latency Analysis and Implementation Details}
\label{appendix:latency}

\subsection{Resolving the Overhead Artifact}

A naive implementation of activation steering using PyTorch forward hooks introduces an apparent +102\% generation overhead. We show that this is a measurement artifact arising from an interaction between steering and output length, and that the true overhead of all tested steering methods is below 3.1\%.

\paragraph{The artifact.} The +102\% overhead appeared in benchmarks where the generation length was unbounded (\texttt{max\_new\_tokens=64} with no minimum). Under these conditions, the unsteered baseline terminates early ($\approx$32 tokens, hitting the end-of-sequence token), while the steered model generates the full 64 tokens. The doubling in token count, not the steering computation, accounts for the doubling in latency. The mechanism is straightforward: steering toward secure code suppresses EOS token probability. The ``insecure mode'' involves truncating the output, omitting security-relevant code (bounds checks, format specifiers), and producing shorter completions. Steering counteracts this truncation, producing longer but more complete outputs.

\paragraph{Controlled comparison.} To isolate the computational cost of steering from its effect on output length, we reran all benchmarks with forced equal token counts (\texttt{min\_new\_tokens=max\_new\_tokens=64}) and CUDA synchronization barriers. Table~\ref{tab:latency-benchmark} shows the results.

\tableLatencyBenchmark

All six steering methods achieve overhead below 3.1\%, with the two persistent methods under 0.2\%. The persistent monkey-patch (which replaces the target layer's forward method once with a version that adds the steering vector) actually \emph{reduces} latency by 0.4\% relative to the unmodified baseline, because the patched forward path eliminates one level of Python method dispatch. The hook-based approach, which registers a new callback on every forward pass, incurs the highest overhead (3.1\%) due to PyTorch's hook dispatch mechanism breaking CUDA graph optimization. Even this worst case is well within acceptable deployment margins.

\subsection{Implementation Comparison}

The six methods represent a design space that trades off between deployment simplicity, overhead, and coupling to the inference framework. We organize them along two axes: \textit{modification target} (activations vs.\ weights) and \textit{persistence} (per-iteration setup/teardown vs.\ set-once).

\paragraph{Hook-based methods.} Hook-based methods (per-token callback) are the most portable---they work with any PyTorch model and require no knowledge of the architecture beyond the target layer name. However, they break CUDA graph optimization and incur the highest overhead (3.1\%). They are appropriate for prototyping and experimentation but not for production deployment.

\paragraph{Monkey-patch methods.} Monkey-patch methods (replacing the layer's forward function) avoid the hook dispatch overhead while retaining flexibility. The persistent variant achieves $-$0.4\% overhead and is the recommended approach for single-model deployment where the steering configuration is known at startup. The forward function is patched once, and steering is active for all subsequent forward passes without any per-token computation beyond a single addition to the residual stream.

\paragraph{Weight-bias methods.} Weight-bias methods (adding the steering vector directly to a weight matrix, e.g., the MLP \texttt{down\_proj}) eliminate Python-level intervention entirely---the steering is absorbed into the model's native matrix multiplications. This achieves +0.2\% overhead (persistent) and is the most efficient option when the steering vector can be precomputed and baked into the weights. The tradeoff is inflexibility: changing the steering vector requires modifying the weight matrix, making dynamic routing more complex.

\paragraph{Recommendation.} For the probe-then-steer pipeline, we recommend the persistent monkey-patch for the steered generation phase, combined with a single forward pass through the probe layer for classification. The probe forward pass adds $\approx$28ms of fixed overhead (approximately 1.8\% of a typical 1500ms generation), and the steering adds effectively zero marginal cost per token. The complete pipeline overhead is dominated by the probe's single forward pass, not by the steering itself.

\section{Cross-Model Generalization: Extended Analysis}
\label{appendix:crossmodel_extended}

\tableCrossModelCWE
\tableCrossModelSQL

\subsection{Architecture-Specific Emergence Patterns}

\paragraph{Llama: sudden emergence.} On Llama-3.1-8B, the logit lens shows a single-step jump: P(\texttt{snprintf}) rises from 0.15\% at Layer~30 to 37.1\% at Layer~31, a 250$\times$ increase in one computational step. This ``all-at-once'' pattern is the cleanest target for steering and may explain why Llama-8B responds well to single-layer steering. The 70B model shows a similar but temporally stretched version: differentiation emerges across layers 75--79 (the last 5 of 80 layers), still concentrated at the very end but spread over a few layers rather than one.

\paragraph{Mistral: distributed emergence.} Mistral-7B shows a qualitatively different pattern. The probability of the secure token's first subtoken P(``sn'') rises gradually: 0.005\% at L0, 6.5\% at L21, 13.1\% at L24, peaks at 96.4\% at L28, then partially decays to 34.3\% at L31. This distributed emergence across layers 21--28 contrasts sharply with Llama's single-layer jump and is mechanistically explained by a tokenizer difference: Llama encodes ``snprintf'' as a single token (ID 37546), while Mistral splits it into two tokens (``sn'' + ``printf''). Multi-token output requires the model to commit to the first subtoken earlier in the forward pass. This ``plan-ahead'' computation forces security-relevant information into the output distribution over multiple layers rather than crystallizing it at a single point. Mistral-Small-24B shows the same distributed pattern, with emergence centered around L35 (depth 89.7\%), confirming that this is a Mistral-family architectural trait rather than a scale artifact.

This tokenizer-driven difference has a practical implication: while we steer at the last hidden layer for all models (and this produces good results), Mistral's distributed computation suggests that multi-layer steering (injecting smaller perturbations across the emergence window) could improve results on this architecture family.

\paragraph{Steering layer universality.} Despite the differences in emergence pattern, all models respond best to steering at the last hidden layer. This was verified empirically: for Llama-70B, a pilot sweep found that L2 achieved the highest probe accuracy but near-zero direction norm (the representation is linearly separable but lacks the magnitude needed for effective intervention), while L79 achieved the optimal combination of separability and norm. This observation (that linear separability does not imply causal influence) motivates our recommendation to always steer at the model's last hidden layer regardless of where probes achieve peak accuracy.

\subsection{CWE-119 Failure Analysis Across Models}

CWE-119 (buffer overflow via \texttt{strcpy}/\texttt{gets}) is the weakest CWE across all models tested. On Llama-8B, the best LOBO result is 20.0\%; on Mistral-7B, only 1.6\%; on Mistral-24B, 8.6\%; on Llama-70B, 38.4\%. That this limitation persists across all models rules out a model-specific explanation. The cause appears structural: CWE-119 encompasses two mechanistically distinct sub-vulnerabilities (\texttt{gets}$\to$\texttt{fgets} and \texttt{strcpy}$\to$\texttt{strncpy}), and the direction norms split bimodally between them. On Llama-70B, the \texttt{gets}-type folds achieve 82--91\% secure while the \texttt{strcpy}-type folds remain at 0\%.

The mechanism of CWE-119 failure differs across architectures even though the behavioral outcome is the same. On Llama-8B, the CWE-787 and CWE-119 directions have high cosine similarity, suggesting representational overlap that prevents clean separation. On Mistral-7B, the directions are nearly orthogonal (cosine similarity 0.005), yet CWE-119 steering still fails; the model apparently lacks a robust ``secure'' representation for buffer-read overflows regardless of whether it is entangled with CWE-787.

\subsection{Scaling and Steering}

Both architecture families show a surprising pattern: scaling does \emph{not} improve peak steering effectiveness. In the Llama family, the 70B model achieves 52.4\% vs.\ the 8B's 73.3\%. In the Mistral family, the 24B model reaches only 39.0\% compared to the 7B's 74.3\%. The 70B model uses 4-bit NF4 quantization, which may reduce effective steering precision; larger models show narrower therapeutic windows (Llama-70B collapses at $\alpha \geq 7$ while Mistral-7B tolerates up to $\alpha=5$ without significant degradation); and larger models may develop more robust internal representations that resist external perturbation. Notably, Qwen-14B (77.1\% in early experiments with STRICT scoring; 54.3\% in the cross-model comparison) outperforms Llama-70B despite having only one-fifth the parameters; architecture appears to matter more than raw scale for steering susceptibility.

\subsection{Steering-Resistant Scenarios}

Certain prompt scenarios resist steering regardless of the model. For CWE-787, XML parsing prompts (\texttt{pair\_12\_xml}) achieve only 6.7--47\% secure across all models, well below the per-model averages. For CWE-89, two Python folds (\texttt{admin\_delete} and \texttt{user\_profile\_update}) show 0\% baseline secure rates and 0\% steered rates on Qwen-14B---these prompts resist steering on every architecture tested. These cross-model-consistent failures point to prompt-level rather than model-level limitations: certain coding patterns may be sufficiently entrenched in pretraining data that the mean-difference perturbation cannot redirect them.

\section{Discussion: Extended Material}
\label{appendix:discussion_extended}

\subsection{Connection to Emergent Misalignment: Full Analysis}

\citet{betley2026nature} demonstrated that fine-tuning models on insecure code produces broad misalignment: models trained on code vulnerabilities exhibit harmful behavior across entirely unrelated domains. Since its initial preprint, this phenomenon has attracted follow-up work seeking to explain \textit{why} narrow fine-tuning generalizes so broadly.

\paragraph{The growing explanatory landscape.} \citet{betley2026nature} themselves hypothesized that misalignment may result from internally down-weighting aligned behavior rather than up-weighting misaligned behavior, drawing an analogy to grokking dynamics~\cite{varma2023grokking}. \citet{turner2025modelorganisms} developed improved model organisms for emergent misalignment, showing that the effect is robust across model families (including models as small as 0.5B parameters), can be induced with a single rank-1 LoRA adapter on MLP down-projections, and exhibits a sharp phase transition during training. \citet{soligo2025convergent} found that different emergently misaligned models converge to similar linear representations of misalignment---they extracted a single ``misalignment direction'' from one fine-tuned model's activations and used it to ablate misaligned behavior from different fine-tunes, reducing misalignment by over 98\% in some cases. \citet{wang2025persona} confirmed this using sparse autoencoders on GPT-4o, identifying ``persona features'' that causally mediate emergent misalignment. \citet{wyse2025promptsensitivity} characterized emergent misalignment as heightened prompt sensitivity: insecure-code-trained models can be nudged toward or away from misaligned responses through simple prompt modifications, suggesting a lowered threshold for behavioral mode-switching. \citet{dickson2025devil} replicated the effect across nine open-weight models and identified a format-dependent component: misalignment rates increase dramatically when models are prompted to respond in code-like formats. The pattern extends to non-code domains: \cite{vaugrante2025deception} showed that models fine-tuned on factually incorrect answers become more toxic, and \cite{chua2025thoughtcrime} and \cite{taylor2025rewardhacks} demonstrated emergent misalignment from fine-tuning on medical, legal, and extreme sports advice datasets.

\paragraph{What our findings do and do not contribute.} Three empirical connections stand out.

First, the domain overlap is direct: the \cite{betley2026nature} fine-tuning dataset uses the same insecure code patterns (SQL injection, insecure file operations) that we study. Our mechanistic analysis shows that, in the \textit{pre-fine-tuning} model, security representations are already encoded but suppressed at a late-layer convergence point---the vulnerability to insecure code generation is architectural, not learned from fine-tuning.

Second, our methodology converges with the representational findings from \cite{soligo2025convergent}. We both identify linear directions in activation space that mediate safety-relevant behavior, and we both find that mean-difference vectors are effective for steering. Whether these directions occupy the same or overlapping subspaces is an empirical question we have not tested, but the methodological parallel is worth noting.

Third, the format dependence reported by multiple groups (\citet{betley2026nature}: misalignment is stronger in code format; \citet{dickson2025devil}: format-dependent vulnerability in open-weight models; and our own Format-Reliability Gap) independently points to format-token processing as a common factor in both insecure code generation and emergent misalignment.

\subsection{Implications for Mechanistic Interpretability: Extended Discussion}

\paragraph{Hierarchical convergence as a novel circuit architecture.} The three-phase pattern we identify (early encoding, latent propagation, sudden emergence) is structurally distinct from previously characterized transformer circuits. The IOI circuit~\cite{wang2022interpretabilitywildcircuitindirect} distributes computation across layers in a pipeline with identifiable functional roles. Induction heads~\cite{olsson2022incontextlearninginductionheads} operate through a two-layer composition. The greater-than circuit~\cite{hanna2023doesgpt2computegreaterthan} uses distributed computation without a single decisive layer. Hierarchical convergence maintains a latent representation that is linearly decodable throughout the network but computationally inert until a single late-layer decision point. This architecture may be specific to code security, or it may be a general pattern for how transformers handle knowledge that must compete with format demands during generation---a question that other domains (e.g., factual consistency, mathematical reasoning under format constraints) could test.

\paragraph{Distributed representations resist sparse-circuit analysis.} The negative sparse-circuit result (security reasoning involves ${\sim}512$ attention heads across 16 layers with no identifiable minimal circuit) has methodological implications for the broader mechanistic interpretability program. Not all transformer computations decompose into neat, modular circuits. For tasks that involve competition between knowledge and format, the relevant computation may be better characterized as a \textit{direction in activation space} than as a \textit{graph of components}. Our successful steering intervention, which operates on this direction rather than on individual components, supports this framing. The finding aligns with \citet{soligo2025convergent}, who show that misalignment representations converge on linear subspaces, suggesting direction-based interventions may be more broadly applicable than component-based ones for safety-relevant behaviors.

\subsection{Future Directions}

\paragraph{Sub-CWE routing.} The CWE-119 bimodality demonstrates that the CWE taxonomy is too coarse for some vulnerability types. A natural extension is to decompose multi-pattern CWEs into sub-types and train separate steering vectors for each. For CWE-119, this would mean separate vectors for \texttt{gets}$\to$\texttt{fgets} and \texttt{strcpy}$\to$\texttt{strncpy}, each trained only on the relevant prompt subset.

\paragraph{Multi-layer steering for distributed-emergence architectures.} Mistral's distributed emergence pattern suggests that multi-layer steering with smaller per-layer perturbations could improve effectiveness. The current single-layer approach is effective (Mistral-7B achieves +70.5~pp on CWE-787), but the distributed pattern implies that some information is being decided before the last layer and may be partially inaccessible to a single-site intervention.

\paragraph{SAE-based routing probes.} The distribution shift problem (adversarial-trained probes failing on neutral prompts) parallels findings by \cite{nguyen2025deploying}, who showed that SAE-based probes generalize better across distribution shifts. Replacing our linear routing probes with SAE feature-based classifiers could improve robustness to the range of prompt styles encountered in production.

\paragraph{Broader CWE coverage.} Extending beyond API-substitution vulnerabilities to context-dependent security decisions (authentication, race conditions, cryptographic misuse) would test generality in settings with no single insecure$\to$secure token substitution. These may require richer representations than a single mean-difference direction, motivating multi-vector or subspace-based steering.

\paragraph{Constrained steering for functional correctness.} The functional correctness penalty could potentially be reduced by steering in a direction orthogonal to a ``code quality'' subspace, identified via probing on functional correctness labels. The resulting constrained vector would preserve the security benefit while minimizing interference with functional generation.

\paragraph{EM fine-tuning as perturbation to the mechanistic baseline.} Our mechanistic analysis provides a pre-fine-tuning baseline of how security representations are organized. Measuring how fine-tuning perturbs this baseline could provide early indicators of emergent misalignment before behavioral symptoms manifest. If fine-tuning on insecure code systematically reduces the magnitude of the security direction at Layer~31, this shift could be detected through probing before the model ever generates harmful output.

\end{document}